%% file: main.tex
\documentclass[10pt,conference]{IEEEtran}
\IEEEoverridecommandlockouts

\usepackage{float}
\usepackage{tabularx}
\usepackage{amsmath}
\usepackage{amsfonts}
\usepackage{algorithmic}
\usepackage[ruled,vlined, linesnumbered]{algorithm2e}
\usepackage{multirow}
\usepackage{graphicx}
\usepackage[export]{adjustbox}
\usepackage{caption}
\usepackage{nccmath}
\usepackage[strict]{changepage}
\usepackage{rotating}
\usepackage[labelformat=simple]{subcaption}

\usepackage{array,booktabs}
\usepackage{algorithm2e}
\usepackage{makecell}
\usepackage{tikz}
\usepackage{ctable}
\usepackage{tablefootnote}
\usepackage{threeparttable}
\usepackage{tcolorbox}
\usepackage{amssymb}
\usepackage{hyperref}

\newcommand{\PreserveBackslash}[1]{\let\temp=\\#1\let\\=\temp}
\newcolumntype{C}[1]{>{\PreserveBackslash\centering}m{#1}}
\newcolumntype{R}[1]{>{\PreserveBackslash\raggedleft}m{#1}}
\newcolumntype{L}[1]{>{\PreserveBackslash\raggedright}m{#1}}

\def\BibTeX{{\rm B\kern-.05em{\sc i\kern-.025em b}\kern-.08em
    T\kern-.1667em\lower.7ex\hbox{E}\kern-.125emX}}
\begin{document}

\title{\huge DDPT: Diffusion-Driven Prompt Tuning for Large Language Model Code Generation  }

\author{\IEEEauthorblockN{Jinyang Li}
\IEEEauthorblockA{\textit{University of Adelaide}\\
Adelaide, SA, Australia\\
jinyang.li01@student.adelaide.edu.au}
\and
\IEEEauthorblockN{Sangwon Hyun}
\IEEEauthorblockA{\textit{University of Adelaide}\\
Adelaide, SA, Australia\\
sangwon.hyun@adelaide.edu.au}
\and
\IEEEauthorblockN{M. Ali Babar}
\IEEEauthorblockA{\textit{University of Adelaide}\\
Adelaide, SA, Australia\\
ali.babar@adelaide.edu.au}
}

\maketitle

\input{00_Abstract.tex}


    \input{01_Introduction}

    \input{02_Background.tex}

    \input{03_Approach.tex}

    \input{04_Experiment.tex}

    \input{05_Futurework.tex}

    \input{07_Conclusion.tex}
    \input{09_acknowledgement.tex}

\input{08_Appendix.tex}
\bibliographystyle{IEEEtran}
\bibliography{sample-base}

\end{document}

%% file: 00_Abstract.tex
\begin{abstract}
 Large Language Models (LLMs) have demonstrated remarkable capabilities in code generation. However, the quality of the generated code is heavily dependent on the structure and composition of the prompts used. Crafting high-quality prompts is a challenging task that requires significant knowledge and skills of prompt engineering. To advance the automation support for the prompt engineering for LLM-based code generation, we propose a novel solution \textbf{Diffusion-Driven Prompt Tuning (DDPT)} that learns how to generate optimal prompt embedding from Gaussian Noise to automate the prompt engineering for code generation. We evaluate the feasibility of diffusion-based optimization and abstract the optimal prompt embedding as a directional vector toward the optimal embedding. We use the code generation loss given by the LLMs to help the diffusion model capture the distribution of optimal prompt embedding during training. The trained diffusion model can build a path from the noise distribution to the optimal distribution at the sampling phrase, the evaluation result demonstrates that DDPT helps improve the prompt optimization for code generation. 
\end{abstract}

%% file: 01_Introduction.tex
\section{Introduction} \label{sec.intro}
 

\begin{figure*}
    \centering
    \includegraphics[width=1\linewidth,height=10.82cm]{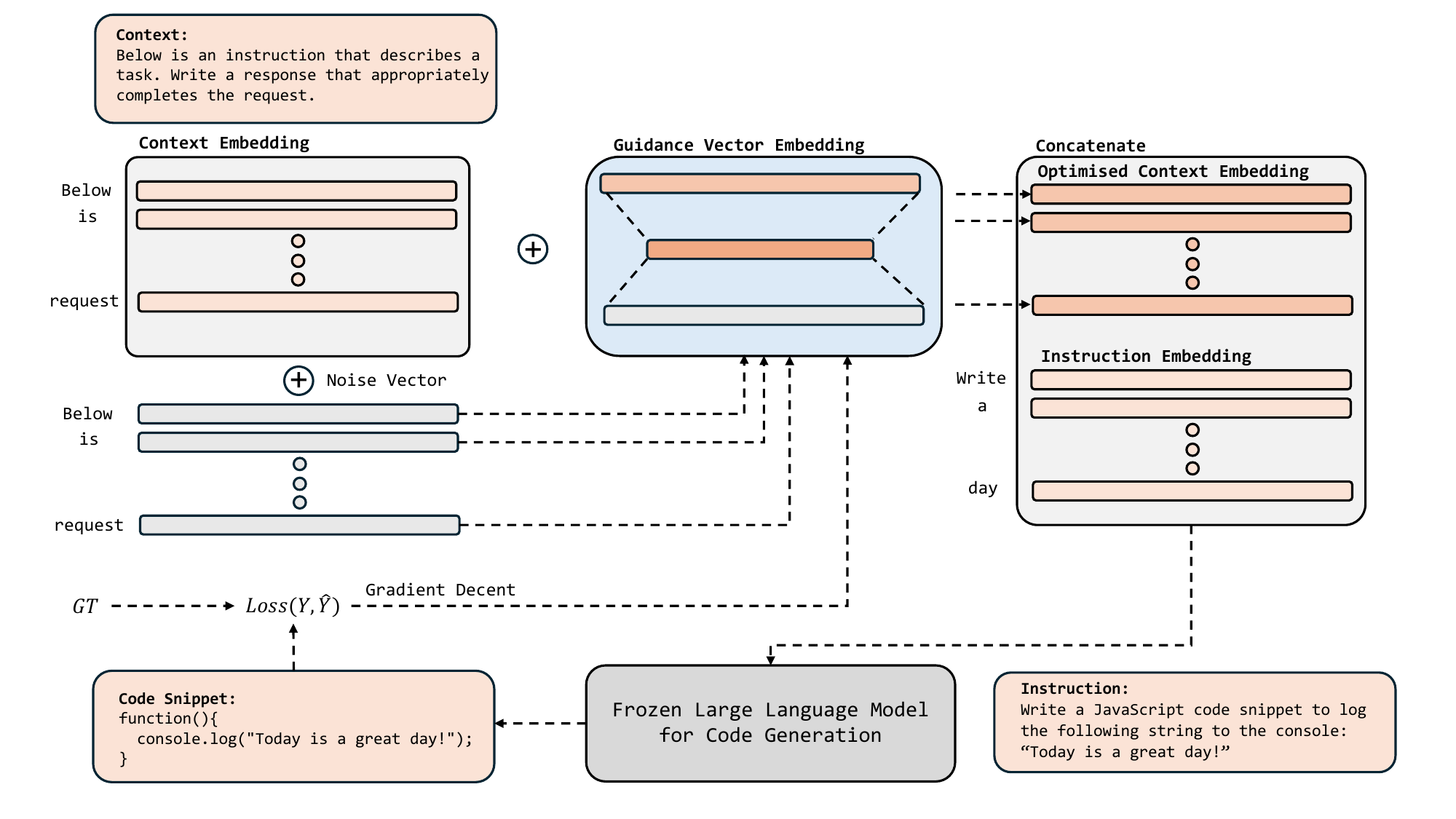}
    \caption{A diagram showing our diffusion-based training process. We first split each input sentence into context and instructions and transform them into word vector embeddings before feeding them into the diffusion model. Gray-colored components indicate Gaussian noise elements and the frozen LLM. GT stands for Ground Truth and text elements that are discrete are highlighted in pink.}
    \label{fig:TrainingProcess}
\end{figure*}

Large Language Models (LLMs), also called foundation models, like ChatGPT~\cite{achiam2023gpt} have garnered significant attention due to their remarkable ability to comprehend human language and generate text outputs accordingly. Based on their large-scale parameter sizes and a broad range of training datasets, several efforts have shown that LLMs can also be successfully applied code generation tasks ~\cite{jiang2024survey}. Using NL2Code (natural-language-to-code)~\cite{jiang2024survey} configuration, a code generation model takes natural language instruction as input and outputs code snippets. However, due to the large sizes of the foundation models and the the context window size limitation ~\cite{holt2023l2mac}, that limits the number of tokens LLM can proceed, instructing a model to perform accurately is considered a challenge.
While it is possible to fine-tune a model using various datasets, the associated computational overhead can be a significant obstacle, particularly with large size LLMs.

Prompt engineering, which focuses on improving the quality of prompt structure, template, and composition, has emerged as a promising solution~\cite{liu2023pre} to obtain the desired output from a LLM. The fundamental premise of prompt engineering is to retrieve the learned knowledge embedding within a language model through the optimization of prompt structure, template, and token selection~\cite{liu2023pre,brown2020language}.
In this setting, a language model is kept `frozen` and is often referred to as a Pre-trained Language Model (PTM) to avoid the huge computational overhead caused by fine-tuning the modern LLMs.  

However, prompt engineering is still a time-consuming process. Significant manual labor efforts are needed to explore the optimal structure of the prompts, and manually searching for the most downstream task-related keywords that provide conditioning to the model is a highly tedious process~\cite{liu2023pre}.

Several studies have addressed the difficulty of producing optimized prompts by automating the prompt engineering process at the prompt embedding level. 
For example, Black Box Tuning (BBT)~\cite{sun2022black} uses a derivative-free Evolution algorithm to produce samples of prompt candidates with lower dimensions than the original prompt sample and uses a random projection matrix to project it back to the dimension of the original sample and add them. 
Prompt Tuning~\cite{lester2021power} proposed to prepend a fixed number of update-able tokens to the original prompt, these tokens have associated trainable parameters and can be updated by gradient computed from language modeling losses on the downstream objective. Prefix tuning~\cite{li2021prefix} is similar to prompt tuning whereas the trainable prompt embedding is prepended to every layer of the model instead of only the input prompt embedding. P-tuning~\cite{liu2023gpt} employs a prompt encoder to optimize trainable prompt embeddings, and the insertion of the embedding is not restricted to prepend. 

Many of the above-mentioned approaches optimize prompt embedding by assigning additional parameters to the embedding and the optimal modification is directly performed over the embedding's trainable parameters. Besides that, prompt embedding optimization suffers from finding a suitable prompt initialization. BBT overcomes this issue by treating the optimized embedding as an add-on to the original prompt embedding. Prompt Tuning has settings that initialize the prompt embedding from random or based on words selected from the LLM's vocabulary space. Initializing prompt embedding based on the real word has little or even negative impact on the downstream task performance as the language model size increases Pre-trained Prompt Tuning (PPT)~\cite{gu2021ppt}.

We propose a novel solution, called \textbf{Diffusion-Driven Prompt Tuning (DDPT)} to address the above-mentioned prompt optimization problems. Drawing inspiration from Diffusion's capability to transform noise into high-quality outputs, our approach, DDPT, moves away from maintaining parameters for prompts. We develop a diffusion-based generator that transforms random noise into a meaningful direction vector. Whilst a LLM is kept frozen during training, an input sentence is decomposed into context and instruction components and converted to word vector embeddings. We perturb the context embedding with Gaussian noise before entering the diffusion model that performs information compression via down projection and generates a directional vector by up-projecting the embedded information to the original word embedding space and guiding the original embeddings to an optimal space through vector addition. The optimized context embedding is concatenated with the instruction embedding to create the input prompt for a frozen LLM. The diffusion model parameters are updated through gradient descent based on the LLM’s code generation loss.

We evaluate our approach's efficacy through language model outputs and quantitative metrics. To address the challenge of interpreting optimized prompt embeddings, we identify the top-k nearest words to each generated token, revealing the semantic relationships within the embeddings.

The key contributions of this work are as follows:
\begin{enumerate}
    \item We explore the use of diffusion as a prompt embedding optimizer and proposed \textbf{DDPT} framework as a novel solution that optimizes prompts through prompt embedding generation. 
    \item Our work demonstrates that a diffusion model, trained with language modeling loss, can successfully learn and generate optimal prompt embeddings. Our approach eliminates the need for embedding parameter storage and encompasses effective random initialization by directly modeling the transformation from random noise to target embeddings.
    \item Our experiment's results show that diffusion optimizer can improve the code generation result produced by LLM. The sampled result obtained from sampling indicates that the model is able to capture the distribution of optimal prompt embedding distribution. Therefore, this is a novel contribution to the diffusion-based optimization and text-domain application.
\end{enumerate}

The remainder of this paper is organized as follows: Section 2 introduces the background and related studies for this research, Section 3 explains our method in detail, Section 4 describes the experiment and empirical analysis result, Section 5 evaluates the threads to validity of our study, Section 6 discusses directions for future work and Section 7 concludes the study. Our code implementation can be found on: \href{https://github.com/yourusername/yourrepository}{\textit{https://github.com/OOGZleo/DDPT}}

%% file: 02_Background.tex
\section{Background and related work} \label{sec.background}
\subsection{\textbf{Language Model For Code Generation}}
Large Language Models (LLMs) take  natural language instruction as input and perform conditional probability modeling over each token generation~\cite{bengio2000neural}. The application of LLMs in transforming natural language descriptions into functional code has emerged as a significant advancement, demonstrating exceptional capabilities~\cite{xu2022systematic, alon2020structural,hindle2016naturalness}. These models interpret input in the form of natural language specifications of programming tasks, which may be supplemented with additional programming context~\cite{jiang2024survey}. 

\subsection{\textbf{Prompt Engineering}}
Prompt Engineering is the process of constructing an optimal prompt template function that results in the best performance on the LLM's downstream task~\cite{liu2023pre,schulhoff2024prompt}. There are generally two types of prompt structure cloze~\cite{cui2021template,petroni2019language} and prefix~\cite{li2021prefix,lester2021power}. Cloze prompt shapes are more suitable for tasks that can be solved using masked LLM whereas prefix prompt shapes are more suitable for tasks involving generation. The most natural way to perform prompt engineering is through manual design~\cite{brown2020language,schick2020exploiting}, however, this is a non-trivial task and even an expert can fail~\cite{jiang2020can}. Moreover, several studies have indicated that the optimal prompt design structure might not be human-readable and may deviate from the syntactic order of natural language processing ~\cite{lester2021power,shin2020autoprompt}.

\subsection{\textbf{Auto Prompt Template Learning}}

Automatic template learning aims to address the limitation of human-readable structure and optimization difficulty by adopting an algorithmic approach that utilizes designed objective functions to explore the optimal prompt structure or representation for model comprehension. 

There are two types of prompts, discrete and soft prompts (a.k.a continuous prompt)~\cite{liu2023pre}. Discrete prompts are constructed using concrete vocabulary tokens drawn from the discrete space related to the downstream task or any set of relevant tokens. These prompts are typically optimized by adjusting the token combination or orders. The concern with discrete prompt optimization is that language models are very sensitive to the different words used for combination or change in the token position thus adjusting them could lead to reduced performance~\cite{gu2021ppt}. 

Soft prompts, conversely, are the embedding form of the discrete prompts. Rather than being constrained by the pre-trained language model's parameters, these prompts incorporate their own distinct parameter set that can be independently optimized. They undergo optimization through gradient descent techniques, typically utilizing neural network architectures. However, soft prompts face challenges in transferability since their gradient updates are specifically tailored to and tightly coupled with the particular model on which they were trained.

\subsection{\textbf{Diffusion}}
Diffusion~\cite{ho2020denoising,sohl2015deep,song2019generative,song2020score} is a recently prominent generative AI technology that has proven successful in numerous applications~\cite{dieleman2022continuous,amit2021segdiff,alcaraz2022diffusion,avrahami2022blended}. It is versatile and capable of generating high-quality samples that exhibit desired properties for specific tasks. let us define a sequence of time steps  \(t \in [0, T]\) and the forward process of diffusion~\cite{ho2020denoising} is defined as follows: 
\begin{flalign}
X_t = \sqrt{\overline{\alpha_t}}*X_{t-1} + \sqrt{1-\overline{\alpha_t}} * z\label{eq:ForwardProcess}
\end{flalign}
\(\sqrt{\overline{\alpha_t}}\) is the drift coefficient, where \( \overline{\alpha_t} = 1 - \overline{\beta_t}\) and \(\sqrt{1-\overline{\alpha_t}}\) is the diffusion coefficient. These coefficients can be interpreted as a weighting between the original data distribution \(x_0 \sim p_0 \) and the Gaussian distributed noise \(z \sim N(0,I)\) determined by the time step. In the forward process, the original data sample distribution is perturbed by randomly selecting a time step \(t \sim uniform({1, ... ,T})\) from the defined time sequence and adding the weighted Gaussian noise \(z\). The model is then tasked with predicting the corresponding noise added at the current time step. As the time step increases, the weight of the noise component grows larger, while the weight of the original data sample decreases. Consequently, at the final time step \(T\), the original data sample is completely perturbed into Gaussian noise. This approach enables training the model to segment the noise added at different time scales, thereby understanding the underlying distribution of time-based noise. The reverse process of diffusion utilizes the trained noise predictor to predict the noise added, starting from the final time step \(T\) and subtracting it from the initialized Gaussian noise \(X_T \sim N(0,I)\) according to the following formula:
\begin{flalign}
X_{t-1} = \frac{1}{\sqrt{\alpha_{t}}}(X_{t} - \frac{1-\alpha_{t}}{\sqrt{1-\alpha_{t}}}\epsilon_{\theta}(X_{t},t)) + \sigma_{t} z
\end{flalign}
$X_{t}$ is the noisy sample at timestep $t$, $\epsilon_{\theta}(X_{t},t)$ is the trained diffusion model that predicts the noise added to the original sample during the forward process, $\sigma_{t}$ is the standard deviation of the Gaussian random noise sample $z \in N(0,I)$, the term $\sigma_{t} z$ ensures the stochastic characteristic of the sampling process preventing it from being deterministic. $\frac{1}{\sqrt{\alpha_{t}}}$ and $\frac{1-\alpha_{t}}{\sqrt{1-\alpha_{t}}}$ balances the scale between the denoised sample and the predicted noise. 
After iterating through all time steps, the original data sample distribution at time step \(t_0\) is ideally restored.

Our methodology involves extending the diffusion model's training objective and its forward process. The specific modifications and their implementation details are thoroughly documented in Section 3.

%% file: 03_Approach.tex
\section{Approach}\label{sec.appr}

\begin{figure*}
    \centering
    \includegraphics[width=1\linewidth]{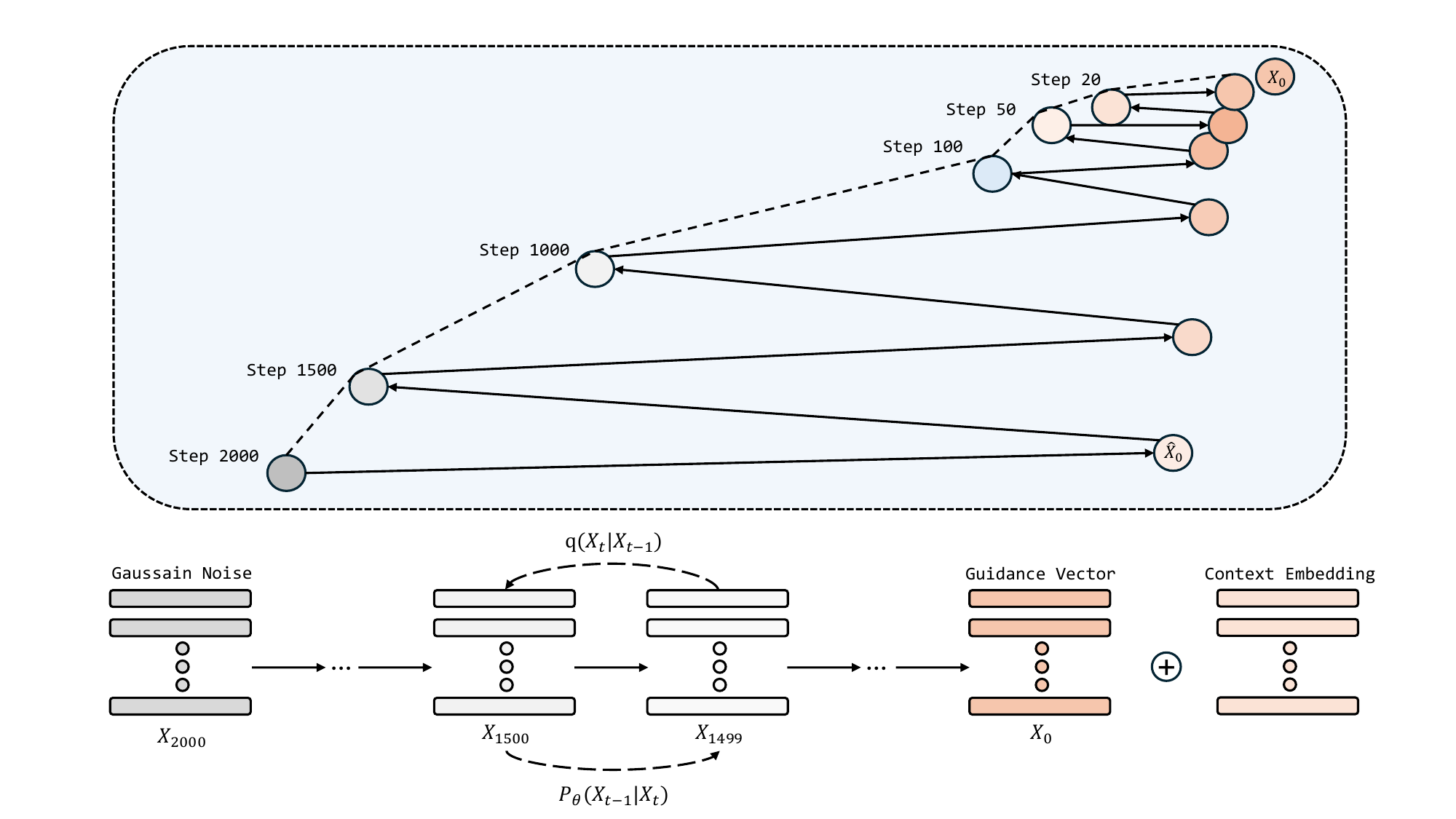}
    \caption{An illustration of our diffusion sampling procedure. The darker components represent the noisy vector embeddings, while the lighter components correspond to the progressively denoised embeddings. The dashed line indicates the modeled transition from the Gaussian noise distribution to the target optimal distribution.}
    \label{fig:SamplingProcess}
\end{figure*}

Our approach optimizes continuous prompt, as we believe the discrete prompt optimization is more computationally challenging~\cite{li2021prefix}. Since the LM (Language Model) attends the embedding vector of each token in the prompt and performs conditional modeling $Pr(Y|n_1,n_2, 
 ... ,n_n)$, we formulate our problem as: Given prompt instruction $P_0 \in \mathbb{R}^{N\times d}$, find the optimal sets of tokens which $P^* = concat(T_1,T_2, ... ,T_N, P_0)$ achieves the best performance on $Pr(Y|P^*)$.
\subsection{\textbf{Preparation}}
Since code generation is a  generative task, the prompt structure in our study is defined as the prefix that follows the same intuition as in ~\cite{lester2021power,li2021prefix}. To allow the language model to generate the instruction tightly, the prompt used in our study requires to have a context part present, and only the context information of a given prompt is treated as prefix and optimized in order to avoid the entire prompt input being rebuilt into global optimal.

To perform optimization over prompt embedding, we first need to convert the discrete prompt sample into a numerical representation that can be understood by the language model. Suppose we have $k$ discrete prompts, each prompt has $n$ context-related tokens and the language model's embedding dimension is 1024, using the language model's embedding table we can convert the $n$ context tokens into prompt embedding of shape $X_{context} \in \mathbb{R}^{n \times 1024}$.

\subsection{\textbf{Optimisation Setting}}
Since the number of dimension space needed to represent the knowledge learned within the LLMs is low~\cite{qin2021exploring,aghajanyan2020intrinsic} which indicates that the effective dimensional space for optimization is lower than the full dimensional space, we follow the intuition mentioned in BBT~\cite{sun2022black} which optimize prompt embedding in a lower dimension space than the original prompt embedding and up project it to the original higher-dimensional space where they act as directional vectors that modify the original embeddings through vector addition.

\subsection{\textbf{Training Objective}}
In order to use diffusion in the optimization setting, we need to perform some modifications over its training objective. Recall the simplified training objective of the diffusion model derived in~\cite{ho2020denoising}:
    \[
        L_{simple}(\theta) := \mathbb{E}_{t,X_{0},\epsilon} [\|X_{0} - X_{\theta}(\sqrt{\overline{\alpha_t}} X_{0} + \sqrt{1-\overline{\alpha_t}} \epsilon, t)\|^{2}]
    \]

Note that we use term $X_{0}$ here rather than noise $\epsilon$ in the original formula as we want the model to focus more on the word embedding itself, the same intuition has also been mentioned in another text-based diffusion study~\cite{li2022diffusion}. 

The training objective is designed to enable the model to recognize patterns in predicting the original sample's distribution from samples with added noise at arbitrary timesteps. The loss calculation, which measures the disparity between predicted and original samples, provides guidance on the accuracy of the model's predictions relative to the ground truth distribution. In our research, we aim to optimize the prompt embedding to reduce the language model's code generation loss. This language modeling loss functions as a directional guide, steering the generation of the directional vector toward the optimal distribution of prompt embeddings. 

We explore a training objective variant and evaluate the performance difference in the next section. This variant challenges the necessity of retaining the noise prediction term in our optimization process, given that both noise prediction loss and language modeling loss guide the model toward particular distributions. Consequently, we eliminate the noise prediction loss and retain only the language modeling loss (Equation~\ref{eq:LMloss}) for our training objective, with y representing the ground truth label and c encompassing all potential labels. Fig~\ref{fig:TrainingProcess} illustrates the training process of our study.

\begin{gather}
L_{LMloss} = -\log \frac{\exp{(X_{y})}}{\sum_{c=1}^{c} \exp{(X_{c})}} \label{eq:LMloss}
\end{gather}
\subsection{\textbf{Sampling}}

We assess the optimization efficiency of diffusion sampling by utilizing the existing DDPM~\cite{ho2020denoising} study's sampling methodology, with several adaptations to the training procedure. Our optimization framework employs a diffusion model that processes only the static contextual information of the prompt, generating an output that represents an optimal prompt distribution not found in the existing dataset. This output serves as input for subsequent sampling iterations. To enhance the model's generalization capabilities and optimization performance, we implemented a novel data augmentation approach. Rather than applying a single perturbation, we utilize the diffusion model's predictions as the base sample and execute the training sequence three additional times within each epoch.

Fig~\ref{fig:SamplingProcess} illustrates the sampling process of our approach. The trained diffusion model generates optimal context embeddings starting from a Gaussian noise vector at the final timestep (t=2000). At each step, the model predicts the optimal context embedding distribution, which is used to derive the context embedding sample for the previous timestep. This iterative prediction process continues until timestep 0, yielding the final optimal context embedding. The procedure effectively constructs a path from the initial Gaussian noise distribution to the target optimal context distribution. $P_{\theta}(X_{t-1}|X_{t})$ models the probability of getting the sample at the previous timestep given the probability of the sample at the current timestep, and $q(X_{t}|X_{t-1})$ models the probability of the sample at the current timestep given the probability of the sample at the previous timestep. The diffusion model learns the modeling of $P_{\theta}(X_{t-1}|X_{t})$ (in our study, $X_{t-1}$ is obtained through asking diffusion model to directly model $X_{0}$ and calculates the probablity of $q(x_{t-1} | x_t, x_0)$), the modeling of $q(X_{t}|X_{t-1})$ is according to the formula of the forward process (Equation~\ref{eq:ForwardProcess}).

\subsection{\textbf{Interpret the optimal embedding}}
Interpreting the optimized prompt embedding is not a straightforward task as these embeddings may not correspond to any existing words in human language or the model's vocabulary set. To derive meaningful interpretations of these optimized prompt embeddings, we calculate the cosine similarity using Equation (\ref{eq:cos}), which allows us to identify the k-nearest word embeddings within the model's vocabulary collection that most closely align with the optimized embedding.
\begin{gather}
\label{eq:cos}
cos(T_i,W_i) = \frac{\sum_{j}^{n}T_{ij}*W_{ij}}{\|T_i\|_2*\|W_i\|_2}
\end{gather}
Where a prompt embedding is formed of tokens of size $n$ by $d$. Suppose we have prompt $p = concat(T_1,T_2, ... ,T_n)$, each token $T_i$ is a word vector with dimension $d$, we compute the pairwise cosine similarity between each word embedding in the token and word embedding $W_i$ in the corpus.
Algorithm~\ref{alg:train} shows the overall training approach of our method. 

\begin{algorithm}[t]
\caption{Training}
\label{alg:train}
\KwData{$P_0 \sim P, \, \hat{Y} \sim Y^c$}
$P_{context}, P_{sentence} = Split(P_0)$\;
$P_{start} = P_{context}$\;
$L_{LMloss} = 0$\;
\For{i in range(k):}{
    $t \sim uniform({1, ... ,T})$\;
    $P_t = \sqrt{\overline{\alpha_t}} P_{start} + \sqrt{1-\overline{\alpha_t}} \epsilon$\;
    $\hat{P}_{start} = M_\theta(P_t) $\;
    
    $P_{\theta} = \hat{P}_{start} + P_{start}$\;
    $P_{\theta} = Concat(P_{\theta}, P_{sentence})$
    
    $ L_{LMloss} = L_{LMloss} + \|LM(P_{\theta})-\hat{Y}\|_2$\;
    $P_{start} = \hat{P}_{start}$\;
}
$ \nabla_\theta Average(L_{LMloss})$\;
\textbf{until converge}\;
\end{algorithm}

%% file: 04_Experiment.tex
\begin{figure*}  
    \centering
    \includegraphics[width=1\linewidth]{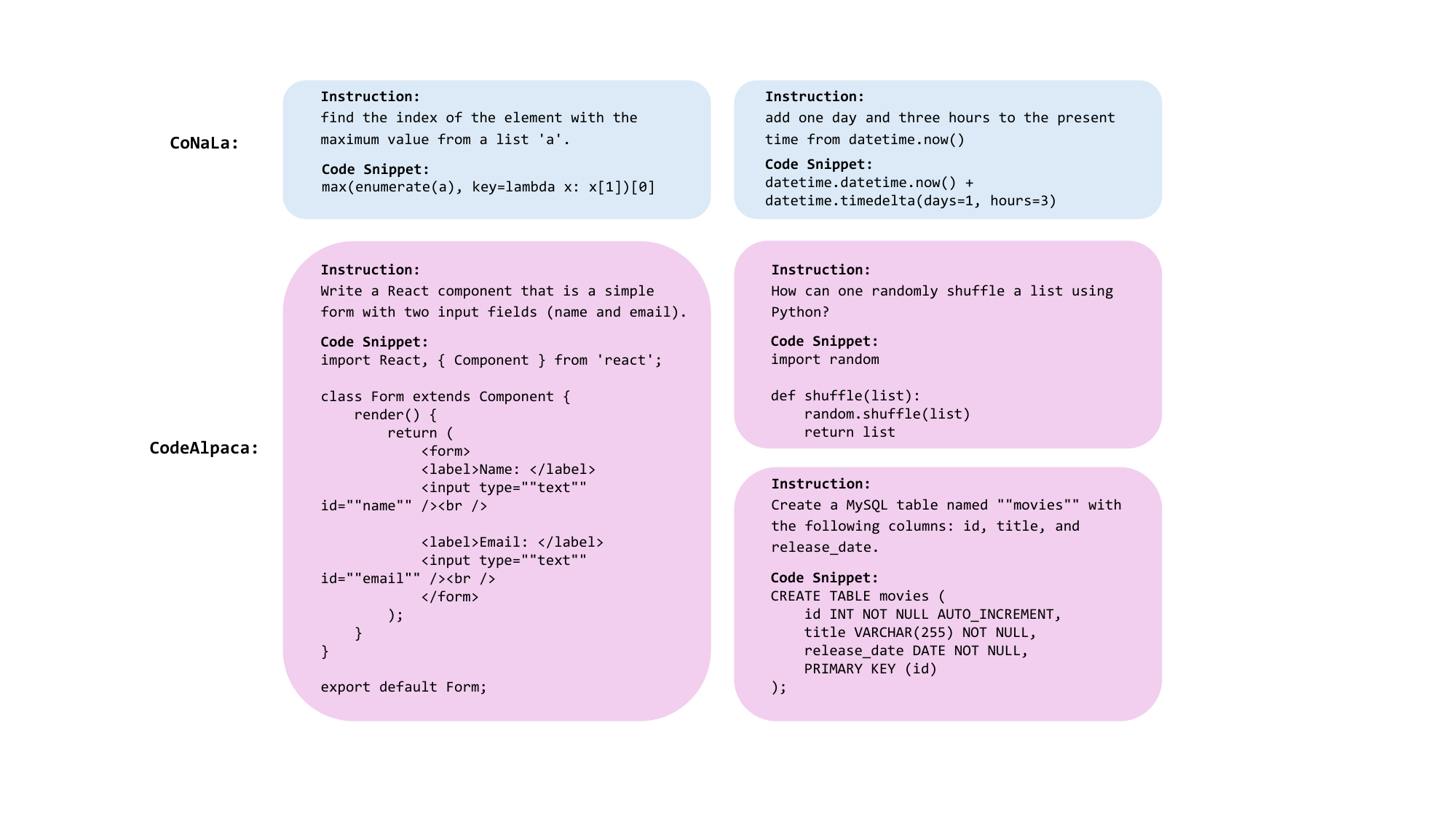}
    \caption{Sample code challenges from the CoNaLa dataset (blue) and CodeAlpaca20k dataset (purple). Though both datasets encompass diverse code generation tasks, they differ in expected output length: CoNaLa typically requires shorter code snippets, while CodeAlpaca20k prompts generally demand longer, more complex code solutions.}
    \label{fig:dataset}
\end{figure*}
\begin{figure}[H]    
    \centering
    \includegraphics[width=0.5\textwidth]{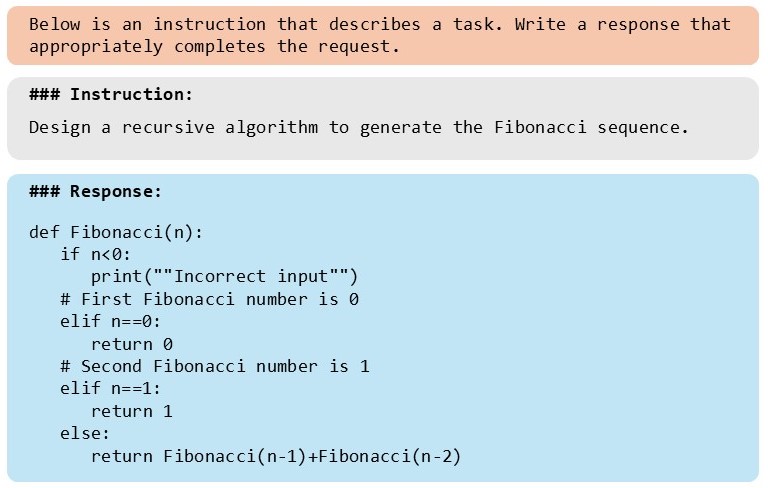}
    \caption{The prompt template used in this study. We uses the default prompt template presented by the codeAlpaca dataset as our initial prompt representation. The template consists of three components: the context portion (orange), the user instruction (gray), and the generated code snippet (blue).}
    \label{fig:PromptTemplate}
\end{figure}

\section{Experiment}
\label{sec.expr}

\subsection{\textbf{Experiment Design}} \label{sec.expr.design}

This section presents our experimental evaluation of the proposed diffusion-based prompt tuning approach. Our method processes the contextual information of a prompt and generates a directional vector that will be added to the original prompt context to enhance the performance of LLMs in code generation tasks. We conduct a thorough analysis of the experimental results, discussing both the strengths and limitations of our approach. Additionally, we outline potential improvements and directions for future research in this area.
The experiments are designed to address the following Research Questions (RQs):

\textbf{RQ1.}
Can diffusion-based prompt tuning enhance the ability of LLMs to generalize across diverse code-generation tasks?
The RQ1 is aimed at examining the effectiveness of our solution in automating prompt engineering and its versatility across a diverse range of programming tasks. It focuses on testing diffusion's ability to capture the distribution of text samples and learning the pattern to predict unseen optimal prompt distribution.

\textbf{RQ2.}
Can the diffusion model improve the quality and correctness of the LLMs code generation through better prompt embedding optimization?
The RQ2 focuses on the code output generated by the LLM using the optimized prompt embedding given by our solution. Specifically, we compare the code result obtained using the original manual prompt input and examine whether the code generated adheres to improved programming language syntax and structure. We also performed a straightforward evaluation of the functional correctness of the generated code snippet. 

\textbf{RQ3.}
Can the optimized prompt embedding be interpreted by humans?
RQ3 aims to address the optimized prompt embedding's interpretability by humans. We are interested in what kinds of information are embedded into these embeddings and can it be understood by humans.

To answer RQ1, we train the diffusion model with code generation datasets that contain various types of programming tasks and evaluate the accuracy of the code generated by the language model using specific metrics. However, to evaluate the performance of LLMs code generation is an area that requires more exploration. In this study, we consider the metrics presented in~\cite{evtikhiev2023out} for evaluation. We use  BLEU-4~\cite{papineni2002bleu} to compare the n-gram matches between generated and reference code, and codeBLEU~\cite{ren2020codebleu} to measure both lexical and syntactic accuracy. We also used METEOR~\cite{banerjee2005meteor} which considers synonyms, word stems, and sequence order, ChrF~\cite{popovic2015chrf} for character-level n-gram precision, and Rouge-L~\cite{lin2004rouge} to find the longest matching sequences. A higher score on any of these metrics indicates better alignment between the generated code and the reference solution. All metrics are applied to both CoNaLa and CodeAlpaca for evaluation except codeBLEU which is only applied to CoNaLa dataset. This is because the codeAlpaca dataset contains programming problems formed with multiple different languages which makes it hard to apply codeBLEU, in opposite CoNaLa dataset only contains Python problems. 

For answering RQ2, we decode the code generation outputs of LLM and perform direct comparison between the result obtained using original prompt and our prompt. We visualize ground truth code snippet as reference in order to evaluate the functional correctness of generated script.

For RQ3, we identify top-k nearest neighbour to the optimized prompt embedding by measuring cosine similarity between them in the vector space. Words that are close in the vector space should have stronger vector projections on each other. We evaluate the category of neighbors to infer the category of the optimized embedding.
\begin{table*}[ht]
\centering
\begin{adjustbox}{width=\textwidth}
\begin{tabular}{lcccccccccccccccc}
\toprule
 & \multicolumn{7}{c}{CodeAlpaca} & \multicolumn{2}{c}{CoNaLa}\\
\cmidrule(lr){3-6} \cmidrule(lr){7-11} 
 & Method & BLEU & ChrF & R-L & MET & BLEU & ChrF & R-L & MET & CodeBLEU\\
 
\midrule
\multirow{2}{*}{\textbf{codeT5p-2B}}
 & Manual Prompt & 13.54 & 22.58 & 25.57 & 26.12 & 2.40 & 13.53 & 12.08 & 11.60 & 2.51\\
 & PT & 14.70 & 24.10 & 26.99 & 27.51 & \textbf{8.87} & \textbf{21.39} & \textbf{25.12} & 13.56 & 5.91\\
 & DDPT & \textbf{16.02} & \textbf{25.82} & \textbf{28.63} & \textbf{30.81} & 6.73 & 17.59 & 17.28 & \textbf{22.96} & \textbf{7.57}\\
\midrule
\multirow{2}{*}{\textbf{codeT5p-6B}}
 & Manual Prompt & 12.07 & 21.18 & 24.20 & 23.93 & 1.81 & 11.08 & 11.11 & 11.53 & 3.15\\
 & PT & 10.15 & 19.55 & 21.90 & 23.17 & 7.35 & \textbf{19.24} & \textbf{21.82} & 11.29 & 5.64\\
 & DDPT & \textbf{14.09} & \textbf{24.26} & \textbf{27.05} & \textbf{28.61} & \textbf{7.49} & 16.96 & 18.81 & \textbf{23.11} & \textbf{8.46}\\
\midrule
\multirow{2}{*}{\textbf{codeT5p-16B}}
 & Manual Prompt & 13.61 & 22.12 & 26.01 & 26.43 & 7.88 & 17.32 & 18.19 & 19.48 & 8.59\\
 & PT & 12.68 & 21.76 & 24.12 & 23.88 & 7.84 & 18.22 & 23.46 & 11.68 & 4.62\\
 & DDPT & \textbf{17.14} & \textbf{26.65} & \textbf{29.58} & \textbf{32.73} & \textbf{14.76} & \textbf{28.59} & \textbf{28.62} & \textbf{31.66} & \textbf{14.50}\\
\midrule
\multirow{2}{*}{\textbf{instructcodeT5p-16B}}
 & Manual Prompt & 16.65 & 29.64 & \textbf{45.81} & \textbf{40.49} & 15.44 & 24.59 & 31.10 & 26.50 & 13.32\\
  & PT & 18.00 & 30.02 & 44.89 & 40.46 & 16.40 & 27.63 & 34.05 & 23.69 & 13.31\\
 & DDPT & \textbf{21.66} & \textbf{31.54} & 37.39 & 38.96 & \textbf{17.02} & \textbf{32.05} & \textbf{34.30} & \textbf{36.09} & \textbf{17.53}\\
\midrule

\bottomrule
\end{tabular}
\end{adjustbox}
\caption{Comparative performance analysis of manual prompts, prompt tuning, and DDPT using CodeT5p models of varying sizes on the CodeAlpaca and CoNaLa datasets, evaluated across multiple metrics.}
\label{tab:experiment}
\end{table*}
\begin{figure*}[ht]
    \centering
    \includegraphics[width=\linewidth,height=9cm]{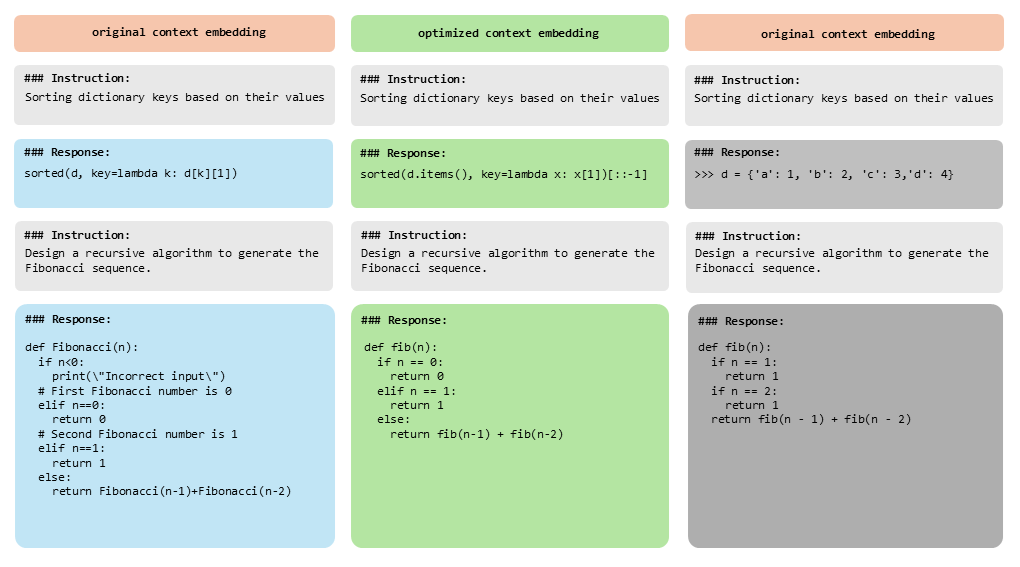}
    \caption{Visual comparison of code generation results across different embedding methods. Orange denotes the original context embedding, green shows the optimized prompt embedding with its generated output, grey indicates code produced using original context embedding, and blue marks the ground truth code snippet. }
    \label{fig:codeCompare}
\end{figure*}
\subsection{\textbf{Dataset}} 
We choose 2 code generation datasets for the case of our study: \textbf{CodeAlpaca}~\cite{chaudhary2023code} is a publicly available dataset that supports multiple programming languages with 20k samples generated by pretrained LLM. Each sample includes instructions, input if any, and a corresponding code snippet. However, the context information for the sample with or without input is different. We conducted our study using only input-free samples, driven by two factors: 1) The computational cost for training with such a large dataset is high, as each diffusion model update required running an inference pass through the language model. 2) Providing proper control to guide diffusion's generation towards specific samples on the text-domain is still an area that is left for exploration. The codeAlpaca dataset provided 9,761 such samples, offering a robust foundation for our training and evaluation objectives.

\textbf{CoNaLa}~\cite{yin2018learning} is a dataset crawled from Stack Overflow and is automatically filtered and manually curated by annotators for evaluating the performance of the system in generating Python code snippets based on natural language description. Both of these datasets contain various programming tasks that support us in answering the research question. Fig~\ref{fig:dataset} demonstrates the variety of programming challenges contained within the CodeAlpaca and CoNaLa datasets.

\subsection{\textbf{Prompt Template Choice}}
To initiate the diffusion training process, a valid prompt embedding sample is required to represent the initial distribution of the original prompt embedding. As previously noted, the model's input is restricted to the context portion of the prompt. Given the challenges in determining a suitable context structure manually, we have adopted the prompt template used in the CodeAlpaca dataset.
Our prompt template is illustrated in Fig~\ref{fig:PromptTemplate}. The structure is as follows: the context information is positioned at the start, the natural language instruction from the dataset is inserted after the instruction header, while the generated code output is positioned following the response header. This standardized format allows for consistency in our approach to prompt engineering and model training.

\subsection{\textbf{Target Models}} 
In this study, We use codeT5p model as our pretrained language model. codeT5p~\cite{wang2023codet5+} model utilizes an encoder-decoder architecture, allowing different types of input to be processed by each component. We fed the context and instruction into the model's encoder and only fed instruction into the decoder as the decoder start ids. This methodology is designed to emulate a prefix language model (prefix-LM)~\cite{tay2022ul2} configuration in order to enhance the fluency of the generated code and ensure that the output closely adheres to the given instructions. Our diffusion model is built on a transformer backbone that mirrors the U-Net design. Like U-Net, it processes input data through a series of down projection and up projection steps, which helps extract relevant features from the input data and match with our optimization setting. Using other backbones is possible but this is out of scope for this study as our work is a pioneer work in exploring the possibility of diffusion-based optimization on the text-domain.

\subsection{\textbf{Experiment Results and Analysis}} \label{sec.expr.result}
\begin{figure*}
    \centering
    \includegraphics[width=\linewidth,height=9cm]{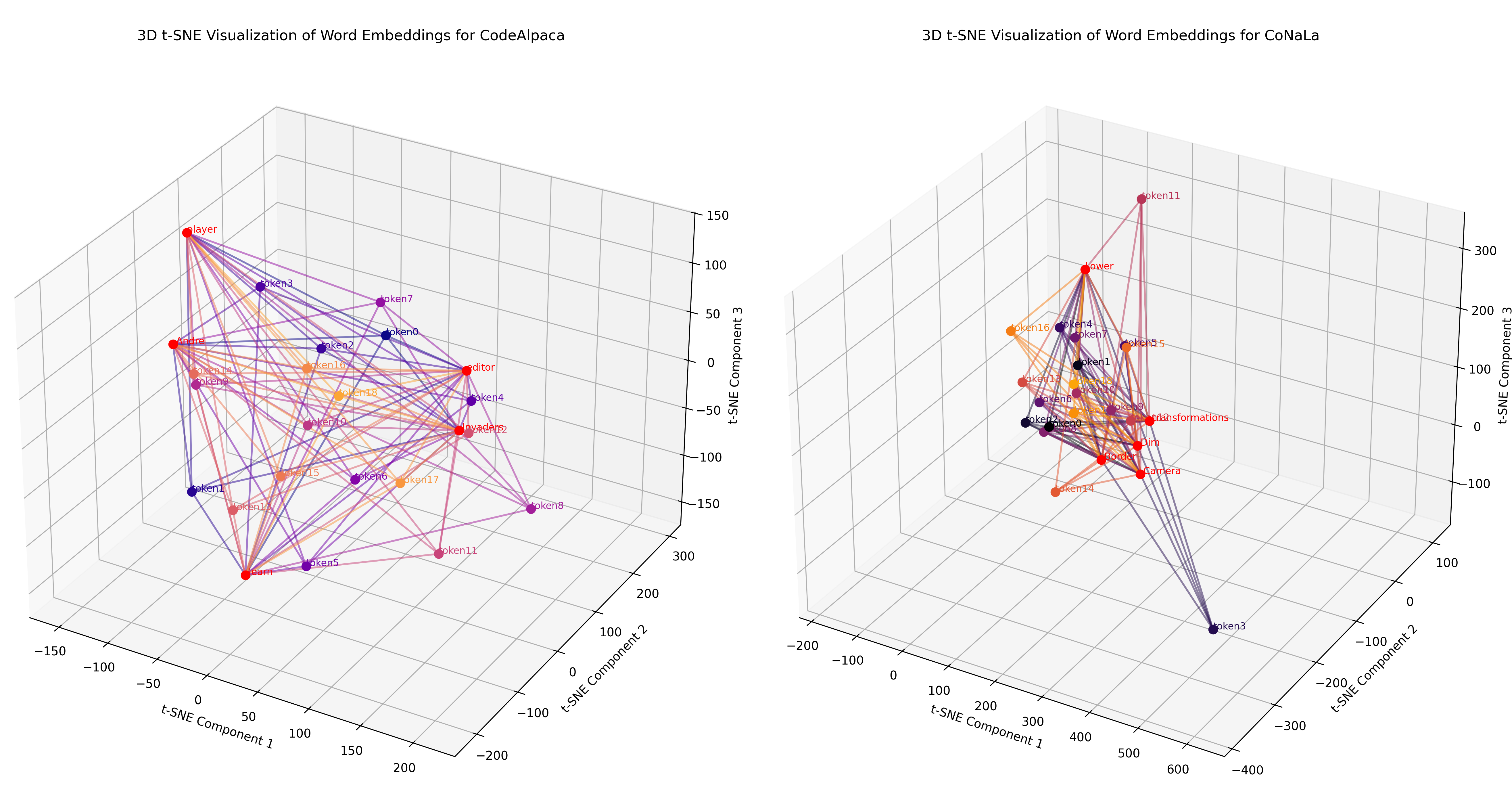}
    \caption{Visualization of top-5 nearest word embeddings for each optimal context token embedding across different datasets. The five nearest neighboring words are highlighted in red, while the generated context tokens are displayed in distinct colors, with connecting lines showing their relationships.}
    \label{fig:topknear}
\end{figure*}
\textbf{RQ1.}
In the experiment for answering RQ1, we train diffusion model with codeT5p series on both CoNaLa and CodeAlpaca dataset. LLMs are set frozen to ensure no gradient update on their parameters. Note that InstructcodeT5p-16B is a finetuned variant of codeT5p series on the CodeAlpaca dataset. Generated code snippets were produced through greedy decoding, and the optimal results were recorded. The evaluation used BLEU-4, ChrF, Rouge-L, METEOR and CodeBLEU metrics, with diffusion sampling set at 2000 sampling time step.  Notably, generating optimal prompt embeddings took under 30 seconds, suggesting text may be easier for diffusion to learn than images.  We refer to the context prompt in the original prompt template as the manual prompt and demonstrate the difference in performance between the optimized prompt and the manual prompt in Table~\ref{tab:experiment}. We notice that Codet5p-16b with our prompt matched or even outperformed finetuned Instructcodet5p-16b's performance across metrics. DDPT also outperforms the manual prompt and prompt-tuning over nearly all metrics for all models. 

On the CodeAlpaca dataset, DDPT showed superior performance for codeT5p models under 16B. For instructcodeT5p-16B model, DDPT showed better scores on BLEU-4 and ChrF, but lower Rouge-L and Meteor scores than manual prompt and prompt-tuning, suggesting DDPT achieves better n-gram matching but worse at code sequence ordering and semantic matching. However, this is only observed on the instructcodeT5p on the CodeAlpaca dataset which might suggest that DDPT forces the model to focus more on the precision of the generated code snippet rather than its learned pattern to focus more on the semantics.

For the CoNaLa dataset, DDPT showed mixed results with CodeT5p-2B and 6B models but significantly better METEOR and CodeBLEU scores. This indicates that DDPT is capable of improving semantic and syntactic understanding of the models. With codeT5p-16B and InstructcodeT5p-16B models, DDPT consistently outperformed other methods across all metrics, demonstrating the diffusion model's effectiveness in optimizing prompt embeddings.

In addition, we notice that manual prompts generally
performed better than prompt-tuning across all models on
CodeAlpaca dataset, with codeT5p-2B and instructcodeT5p-16B being the sole
exception. This pattern wasn’t seen with the CoNaLa dataset.
We believe this difference stems from CodeAlpaca’s higher
complexity and the significant differences in the length of
generated code output as shown in Fig 3. The improvement seen in the codeT5p-2B model is likely due to its smaller parameter size, which simplifies optimization. In the case of the instructcodeT5p-16B model, its enhanced performance with prompt tuning is likely a result of LLM fine-tuning. While alternative decoding strategies like sampling or beam search might enhance prompt tuning results, we limited our decoding method to basic greedy decoding for the purpose of this study.

\textbf{RQ2.}
To address RQ2, we conducted a straightforward analysis manually between code generated using the manual prompt template provided by the CodeAlpaca dataset and our optimized prompts, as illustrated in Fig~\ref{fig:codeCompare}. The figure presents three code snippets: the ground truth (in blue), output from our optimized prompt (in green), and output from the manual prompt template (in gray). In the Fibonacci sequence implementation example, while both approaches produced partially correct solutions, they differ in their handling of base conditions. Our optimized prompt led to the correct logic ("if n==0: return 0 elif n==1: return 1"), whereas the manual prompt resulted in incorrect conditions ("if n==1: return 1, if n==2 return 1"). This demonstrates our model's enhanced ability to guide LLMs toward semantically accurate code generation. Similarly, for the dictionary key sorting task, the manual prompt generated code that adhered more to the literal interpretation of the instruction but failed to meet the intended functionality. In contrast, our optimized prompt produced code that more closely matched the ground truth solution, better fulfilling the user's requirements. 

\textbf{RQ3.}
Addressing RQ3, Fig~\ref{fig:topknear} presents a t-SNE visualization of the five closest word embedding neighbors to our optimized prompt across different datasets. The visualization reduces high-dimensional data to facilitate a better understanding of the embeddings in 3D space. Each of the 19 tokens in the optimised context prompt is assigned distinct colors and shows connections to its five nearest neighboring words. In the CodeAlpaca dataset, the optimized embedding's closest neighbors were "editor", "learn", "player", "invaders", and "Andre" with most terms falling into categories related to actions and roles, particularly those involving modification behaviors. For the CoNaLa dataset, the nearest neighbors were "transformations", "Border", "Camera", "Dim", and "lower" with most terms relating to spatial concepts or adjustment actions. These results indicate that the diffusion model developed the ability to create guiding vectors, which steer the original prompt's embeddings toward action-focused terms that are partially aligned with the semantics of modification in the downstream application domain. While the generated embeddings may not be directly understandable by humans, they can be interpreted by identifying the nearest neighbor words. A notable observation was the tendency of tokens to share the same nearest word embeddings, suggesting that there might be a limit in the number of words in the LLM's collection or DDPT learned to centralize all embeddings towards an optimal area in the embedding space.

%% file: 05_Futurework.tex
\section{Threats to Validity}
This section introduces the threats to validity we faced in our research and outlines how we plan to mitigate these challenges in our future studies. 

\subsection{\textbf{Evaluate our approach against a broader range of prompt-based learning methods and test it across a more diverse selection of language models}}
In this study, we compare our approach with the prompt tuning study~\cite{lester2021power} which is representative of the prompt embedding optimization technique, and update the prompt embedding based on gradient descent. While our method showed better results than traditional prompt tuning and matched the performance of LLM finetuning, more comparison between diverse prompt-based learning techniques such as gradient-free approaches like \textbf{BBT}~\cite{sun2022black} that uses evolution algorithm or reinforcement learning approach like \textbf{RLPrompt}~\cite{deng2022rlprompt} could be conducted. Due to limited computational power, we only use the codet5p series in this study. The unified training objective~\cite{tay2022ul2} allows the codet5p model to handle various natural language processing tasks effectively and reaches comparable performance on natural language to code tasks with state-of-the-art decoder-only models. Our future studies should expand to test our approach with different LLM architectures and analyses with a broader range of prompt-based learning methods when more computational resources are available.

\subsection{\textbf{Security measurement of the generated code snippet and vulnerability analysis}}
In this study, We compared code generated using our prompt optimization approach against code from manual prompts. While our method showed improvements in code quality and accuracy, we haven't yet examined the security aspects of the generated code. Since our research primarily focused on using diffusion for prompt embedding optimization, a thorough analysis of security vulnerabilities will need to be addressed in future work to make this approach more suitable for practical applications. We also need to investigate how different decoding strategies in LLM code generation might affect security, as probability-based token selection could potentially introduce new vulnerabilities.

\subsection{\textbf{Evaluating how various decoding strategies affect language models' ability to generate code}}
Our study used simple greedy decoding for code generation, though this approach isn't always ideal for getting the best results from language models and we observed repeated tokens in the code generation output using the optimized prompt. To address the issue of repeated tokens that we observed during the experiment, we implemented two fixes: a 1.2 repetition penalty and a no-repeat n-gram setting of 2. While these adjustments helped, there's potential for better performance through more sophisticated approaches like beam search or temperature-based sampling, which could both reduce repetition and potentially improve the model's overall output quality.

\subsection{\textbf{Assess the functional correctness of the generated code by measuring its execution performance}}
In this research, we did not employ execution-based metrics such as Pass@k~\cite{chen2021evaluating} to evaluate the functional correctness of the generated code. The primary focus of this study is to enhance the code generation quality of large language models (LLMs) within the NL2Code setting. The prompts used in this study consist of brief and straightforward natural language instructions rather than detailed function signatures or docstrings. Some prompts require the LLM to produce code snippets, such as SQL queries, which necessitate server and database configurations. This presents challenges in developing unit tests due to the significant time and effort required to create custom test cases. We plan to address this in future work by designing more sophisticated unit tests and assessing functional correctness using the Pass@k metric.

\section{Discussion}
This section explores potential improvements to our research outcome by focusing on two main limitations of the methodology: 1) The adaptability of our input handling. 2) Our ability to control the diffusion sampling process.

\subsection{\textbf{Limitation on the length of text input}}
Our current approach has a significant limitation regarding input flexibility. We trained our diffusion model to work with fixed-length prompts by using the context portion of our prompt template. Since this context embedding serves as our original distribution, the model can only generate directional vectors matching this fixed length. This design choice prevents us from investigating how varying prompt lengths might affect code generation quality. To overcome this constraint, future research should explore methods that allow the diffusion model to handle variable-length text samples, which would make our framework more adaptable and comprehensive.

\subsection{\textbf{Optimal prompt embedding generation through controllable diffusion sampling}}

A key limitation of our study is the lack of control over the diffusion sampling process. While we adapted the DDPM~\cite{ho2020denoising} sampling approach, which starts with Gaussian noise at timestep T and generates guidance vectors for prompt optimization, the path from noise to optimal distribution remains as the blackbox. Unlike traditional approaches that predict the original prompt embedding, our model predicts guidance vectors to direct prompt movement. Although controlled sampling has been studied in both vision and language domains, our novel diffusion-based optimization approach means there is no existing research on controlled sampling methods for diffusion optimizers. Future work should investigate ways to make this process more interpretable and controllable, potentially enabling the generation of specialized prompt embeddings for specific code generation task.

%% file: 07_Conclusion.tex
\section{Conclusion}
\label{sec.conclusion}

Our research investigates the application of diffusion-based techniques for generating optimized prompt embeddings in NL2Code generation tasks instead of performing optimization on the parameters of the prompt. We introduce \textbf{DDPT}, an innovative approach that generates a directional vector from Gaussian noise, which, when added to the original prompt embedding, guides it toward an optimal distribution in the embedding space. Experimental findings demonstrate that our methodology effectively improves the quality of LLM-generated code. Through visualization of the k-nearest words to the generated embeddings, we show that our diffusion model is capable of capturing the semantic characteristics of downstream tasks to some degree. The application of diffusion models for optimized sample generation represents a significant advancement in both text-domain diffusion applications and prompt-based learning research.

%% file: 09_acknowledgement.tex
\section{Acknowledgment}
\label{sec.acknowledgment}
I am grateful for the support and guidance offered by the CREST members. Additionally, I would like to extend my special thanks to Ziyang Ye for assisting me with code implementation and engaging in discussions to help me resolve various issues.